\documentstyle[12pt,twoside,fleqn,espcrc1]{article}

\title{ On nature of  scalar $a_0(980)$ and $f_0(980)$-mesons}

\author{N.N. Achasov \address{Laboratory of Theoretical Physics, \\
Sobolev Institute for Mathematics,\\
Academician Koptiug prospekt, 4,\\
630090 Novosibirsk 90, Russia}}

\begin{document}

\maketitle

\begin{abstract}
It is presented a critical consideration of all unusual properties of the
scalar $a_0(980)$ and $f_0(980)$-mesons in the four-quark, two-quark and
molecular models. The arguments are adduced that the four-quark model is more
preferable. It is discussed the complex of experiments that could finally
resolve this issue.
\end{abstract}

\section{Introduction}
Spherical Neutral Detector (SND) from the $e^+e^-$-collider VEPP-2M in
Novosibirsk has discovered \cite{snda098,sndf098} the electric dipole decays
$\phi\to\gamma\pi^0\pi^0$ and $\phi\to\gamma\pi^0\eta$ in the region of the
soft by strong interaction standard photons with the energy
$\omega < 120\,\mbox{MeV}$, i.e. in the region of the scalar $a_0(980)$ and
$f_0(980)$-mesons  $m_{\pi^0\pi^0}> 900\,
\mbox{MeV}$ and $m_{\pi^0\eta}> 900\,\mbox{MeV}$,
$\phi\to\gamma f_0(980)\to\gamma\pi^0\pi^0$ and
$\phi\to\gamma a_0(980)\to\gamma\pi^0\eta$. The data are
\begin{eqnarray}
\label{snd1}
&& B(\phi\to\gamma\pi^0\pi^0\,;\,m_{\pi^0\pi^0}>900\,\mbox{MeV})=
(0.5\pm 0.06\pm 0.06)\cdot 10^{-4}\nonumber\\
&&\mbox{at total}\ \ B(\phi\to\gamma\pi^0\pi^0)=
(1.14\pm 0.10\pm 0.12)\cdot 10^{-4}\,,\\
\label{snd2}
&& B(\phi\to\gamma\pi^0\eta\,;\,m_{\pi^0\eta}>900\,\mbox{MeV})\simeq
0.5\cdot 10^{-4}\,,\nonumber\\
&&\mbox{at total}\ \ B(\phi\to\gamma\pi^0\eta)=(0.83\pm 023)\cdot 10^{-4}\,.
\end{eqnarray}

Cryogenic Magnetic Detector-2 (CMD-2) from the $e^+e^-$-collider
VEPP-2M in Novosibirsk has confirmed both of these results \cite{solodov-99}.

The branching ratios in Eqs. (\ref{snd1}) and (\ref{snd2}) are great for this
photon energy region and, probably,  can be understood only if four-quark
resonances are produced \cite{achasov-89,achasov-97}.
Note that the $a_0(980)$ meson is produced in the $\phi$ radiative
decay as intensively as the containing strange quarks $\eta'$ meson.

\section{Evidences for strange quarks in the $f_0(980)$ and 
$a_0(980)$-mesons}

To feel why numbers in Eqs. (\ref{snd1}) and (\ref{snd2}) are great, one can
adduce the rough estimate. Let there be structural radiation without a
resonance in the final state with the spectrum
$$\frac{d\Gamma(\phi\to\gamma\pi^0\pi^0(\eta))}{d\omega}\sim
\frac{\alpha}{\pi}\cdot\delta_{OZI}\cdot\frac{1}{m^3_{\phi}}\omega^3\,,$$
where $\delta_{OZI}\sim 10^{-2}$ is a factor describing the suppression
by Okubo-Zweig-Iizuka (OZI) rule. Then one gets
$$\Gamma(\phi\to\gamma\pi^0\pi^0(\eta))\sim\frac{1}{4}\cdot\frac{\alpha}{\pi}
\cdot\delta_{OZI}\cdot\frac{\omega^4_0}{m^3_{\phi}}\simeq 10^{-6}\,
\mbox{MeV}\,,$$ 
$$ B(\phi\to\gamma\pi^0\pi^0(\eta))\sim 2\cdot 10^{-7}\,,
\quad\mbox{for}\quad\omega_0=120\,\mbox{MeV.}$$

\subsection{Evidences for strange quarks in the $a_0(980)$-meson}

To understand, why Eq. (\ref{snd2}) points to four-quark model, is particular
easy. Really, the $\phi$-meson is the isoscalar practically pure
$s\bar s$-state, that decays to the isovector hadron state $\pi^0\eta$ and
the isovector photon. The isovector photon originates from the $\rho$-meson,
$\phi\to\rho a_0(980)\to\gamma\pi^0\eta$, the structure of which in this
energy region is familiar
\begin{eqnarray}
\label{rho}
&&\rho\approx (u\bar u-d\bar d)/\sqrt{2}\,.
\end{eqnarray}

The general structure of the $a_0(980)$-meson , from which the 
$\pi^0\eta$-system originates, is
\begin{eqnarray}
\label{a0}
&& a_0(980)=c_1(u\bar u-d\bar d)/\sqrt{2}+
c_2s\bar s(u\bar u-d\bar d)/\sqrt{2} + ...\, .
\end{eqnarray}

The strange quarks, with the first term in Eq. (\ref{a0}) taken as
dominant, are absent in the intermediate state.  So, we would have the
suppressed by OZI-rule decay with
$B(\phi\to\gamma a_0(980)\to\gamma\pi^0\eta)\sim 10^{-6}$ owing to
the real part of the decay amplitude \cite{achasov-97}. The imaginary part
of the decay amplitude, resulted from the $K^+K^-$- intermediate state
($\phi\to\gamma K^+K^-\to\gamma a_0(980)\to\gamma\pi^0\eta$), violates the
OZI-rule and increases the branching ratio \cite{achasov-89,achasov-97} up
to $10^{-5}$.

The four-quark hypothesis is supported also by the $J/\psi$-decays. Really,
\cite{pdg-98}
\begin{eqnarray}
\label{a2rho}
&& B(J/\psi\to a_2(1320)\rho)= (109\pm 22)\cdot 10^{-4}\,,
\end{eqnarray}
while \cite{kopke-89}
\begin{eqnarray}
\label{a0rho}
&& B(J/\psi\to a_0(980)\rho)< 4.4\cdot 10^{-4}\,.
\end{eqnarray}

The suppression
\begin{eqnarray}
\label{a0rho/a2rho}
&& B(J/\psi\to a_0(980)\rho) /B(J/\psi\to a_2(1320)\rho)< 0.04\pm 0.008
\end{eqnarray}
seems strange, if one considers the $a_2(1320)$ and $a_0(980)$-states
as the tensor and scalar two-quark states from the same P-wave multiplet
with the quark structure
\begin{eqnarray}
\label{a0qq}
&& a_0^0=(u\bar u-d\bar d)/\sqrt{2}\ \ ,\ \ a^+_0=u\bar d\ \ ,\  a^-_0
= d\bar u\,.
\end{eqnarray}
While the four-quark nature of the $a_0(980)$-meson with the symbolic
quark structure
\begin{eqnarray}
\label{a0qqqq}
&& a^0_0=s\bar s(u\bar u-d\bar d)/\sqrt{2}\ \ ,\ \ a_0^+=s\bar su\bar d\ \ ,
\ \ a_0^-=s\bar sd\bar u
\end{eqnarray}
is not c®ntrary to the suppression in Eq. (\ref{a0rho/a2rho}).

Besides, it was predicted in \cite{achasov-82} that the production vigor
of the $a_0(980)$-meson, with it taken as the four-quark state from the
lightest nonet of the MIT-bag \cite{jaffe-77}, in the
$\gamma\gamma$-collisions should be suppressed by the value order in
comparison with the $a_0(980)$-meson taken as the two-quark P-wave state.
In the four-quark model there was obtained the estimate  \cite{achasov-82}
\begin{eqnarray}
\label{ga0gg4q}
&& \Gamma(a_0(980)\to\gamma\gamma)\sim 0.27\,\mbox{keV,}
\end{eqnarray}
which was confirmed by experiment \cite{crystalball,jade}
\begin{eqnarray}
\label{ga0ggexp}
&&\Gamma (a_0\to\gamma\gamma)=(0.19\pm 0.07 ^{+0.1}_{-0.07})/B(a_0\to\pi\eta)
\,
\mbox{keV, Crystal Ball,}\nonumber\\
 && \Gamma (a_0\to\gamma\gamma)=(0.28\pm 0.04\pm 0.1)/B(a_0\to\pi\eta)\,
\mbox{keV, JADE.}
\end{eqnarray}
At the same time in the two-quark model (\ref{a0qq}) it was anticipated
\cite{budnev-79,barnes-85} that
\begin{eqnarray}
\label{ga0gg2q}
\Gamma(a_0\to\gamma\gamma)=(1.5 - 5.9)\Gamma (a_2\to\gamma\gamma)=
(1.5 - 5.9)(1.04\pm 0.09)\,\mbox{keV.}
\end{eqnarray}
The wide scatter of the predictions is connected with different reasonable
guesses of the potential form.

\subsection{Evidences for strange quarks in the $f_0(980)$-meson}

As for the $\phi\to \gamma f_0(980)\to\gamma\pi^0\pi^0$-decay, the more
sophisticated analysis is required.

The structure of the $f_0(980)$-meson, from which the $\pi^0\pi^0$-system
originates, in general, is
\begin{eqnarray}
\label{f0}
&& Y=f_0(980)=\tilde c_0gg+\tilde c_1(u\bar u+d\bar d)/\sqrt{2}+
\tilde c_2s\bar s + \tilde c_3s\bar s(u\bar u+d\bar d)/\sqrt{2} + ...\ .
\end{eqnarray}

First we discuss  a possibility to treat the $f_0(980)$-meson as the
quark-antiquark state.

The hypothesis that the $f_0(980)$-meson is the lowest two-quark P-wave
scalar state with the quark structure
\begin{eqnarray}
\label{f0qq}
&& f_0=(u\bar u+d\bar d)/\sqrt{2}
\end{eqnarray}
contradicts Eq. (\ref{snd1}) in view of OZI, much as Eq. (\ref{a0qq})
contradicts Eq. (\ref{snd2}) (see  the above arguments).

Besides, this hypothesis contradicts a variety of facts:\\
i) the strong coupling with the $K\bar K$-channel
\cite{achasov-84,achasov-97}
\begin{eqnarray}
\label{r}
1<R=|g_{f_0K^+K^-}/g_{f_0\pi^+\pi^-}| ^2\leq 8\,,
\end{eqnarray}
for from Eq. (\ref{f0qq}) it follows that
$|g_{f_0K^+K^-}/g_{f_0\pi^+\pi^-}|^2=\lambda/4\simeq 1/8$, where $\lambda$
takes into account the strange sea suppression;\\
ii) the weak coupling with gluons \cite{eigen-88}
\begin{eqnarray}
\label{f0gluons}
&& B(J/\psi\to\gamma f_0(980)\to\gamma\pi\pi) < 1.4\cdot 10^{-5}
\end{eqnarray}
opposite the expected one \cite{farrar-94} for Eq. (\ref{f0qq})
\begin{eqnarray}
\label{farrar2q}
&& B(J/\psi\to\gamma f_0(980))\geq B(J/\psi\to\gamma f_2(1270))/4\simeq
3.4\cdot 10^{-4}\,;
\end{eqnarray}
iii) the weak coupling with photons \cite{marsiske-90,gidal-88}
\begin{eqnarray}
\label{gf0ggexp}
&&\Gamma (f_0\to\gamma\gamma)=(0.31\pm 0.14\pm 0.09)\,
\mbox{keV, Crystal Ball,}\nonumber\\
&& \Gamma (f_0\to\gamma\gamma)=(0.24\pm 0.06\pm 0.15)\, \mbox{keV, MARK II}
\end{eqnarray}
opposite the expected one \cite{budnev-79,barnes-85} for Eq. (\ref{f0qq})
\begin{eqnarray}
\label{gf0gg2q}
&&\Gamma(f_0\to\gamma\gamma)=(1.7 - 5.5)\Gamma (f_2\to\gamma\gamma)=
(1.7 - 5.5)(2.8\pm 0.4)\,\mbox{keV;}
\end{eqnarray}
iv) the decays $J/\psi\to f_0\omega$, $J/\psi\to f_0\phi$,
$J/\psi\to f_2\omega$, $J/\psi\to f_2'\phi$ \cite{pdg-98}
\begin{eqnarray}
\label{f0omega}
B(J/\psi\to f_0(980)\omega)=(1.4\pm 0.5)\cdot 10^{-4}\,.
\end{eqnarray}
\begin{eqnarray}
\label{f0phi}
B(J/\psi\to f_0(980)\phi)=(3.2\pm 0.9)\cdot 10^{-4}\,.
\end{eqnarray}
\begin{eqnarray}
\label{f2omega}
B(J/\psi\to f_2(1270)\omega)=(4.3\pm 0.6)\cdot 10^{-3}\,,
\end{eqnarray}
\begin{eqnarray}
\label{f2'phi}
B(J/\psi\to f_2'(1525)\phi)=(8\pm 4)\cdot 10^{-4}\,,
\end{eqnarray}

The suppression
\begin{eqnarray}
\label{f0omega/f2omega}
&& B(J/\psi\to f_0(980)\omega) /B(J/\psi\to f_2(1270)\omega)= 0.033\pm 0.013
\end{eqnarray}
looks strange in the model under consideration as well as Eq.
(\ref{a0rho/a2rho}) in the model (\ref{a0qq}).

The existence of the $J/\psi\to f_0(980)\phi$-decay of greater intensity than
the $J/\psi\to f_0(980)\omega$-decay ( compare Eq. (\ref{f0omega}) and Eq.
(\ref{f0phi}) ) shuts down the model (\ref{f0qq}) for in the case under
discussion the $J/\psi\to f_0(980)\phi$-decay should be suppressed in
comparison with the $J/\psi\to f_0(980)\omega$-decay by the OZI-rule.

So, Eq. (\ref{f0qq}) is excluded at a level of physical rigor.

It is impossible to consider the $f_0(980)$-meson as the near $s\bar s$-state
without a gluon component \cite{achasov-98}. 

The introduction of a gluon component, $gg$, in the $f_0(980)$-meson
structure  allows the weak coupling with gluons (\ref{f0gluons}) to be
resolved easy. Really,  by \cite{farrar-94},
\begin{eqnarray}
\label{rtogg}
&& B(R[q\bar q]\to gg)\simeq O(\alpha_s^2)\simeq 0.1 - 0.2\,,\nonumber\\
&& B(R[gg]\to gg)\simeq O(1)\,,
\end{eqnarray}
then the minor ($\sin^2\alpha\leq 0.08$) dopant of the gluonium
\begin{eqnarray}
\label{f0ss}
&& f_0=gg\sin\alpha +\left [\left (1/\sqrt{2}\right )(u\bar u+d\bar d)
\sin\beta + s\bar s\cos\beta\right ]\cos\alpha\,,\nonumber\\
&&\tan\alpha=-O(\alpha_s)\left (\sqrt{2}\sin\beta +\cos\beta\right )\,,
\end{eqnarray}
allows to satisfy Eqs. (\ref{r}), (\ref{f0gluons}) and to get the weak
coupling with photons
\begin{eqnarray}
\label{f0togamagammanearss}
&& \Gamma (f_0(980))\to\gamma\gamma)< 0.22\,\mbox{keV}
\end{eqnarray}
at
\begin{eqnarray}
\label{tgbeta}
&& -0.22>\tan\beta > -0.52\,.
\end{eqnarray}

So, $\cos^2\beta > 0.8$ and the $f_0(980)$-meson is near the $s\bar s$-state,
as in \cite{nils-82}.

The scenario, in which with Eq. (\ref{f0ss}) the $a_0(980)$-meson is
the two-quark state (\ref{a0qq}), runs into following difficulties:\\
i) it is impossible to explain the $f_0$ and $a_0$-meson mass degeneration;\\
ii) it is possible to get only
\cite{achasov-89,achasov-97}
\begin{eqnarray}
\label{phigammaf0a0}
&& B(\phi\to\gamma f_0\to\gamma\pi^0\pi^0)\simeq 1.7\cdot 10^{-5}\,, \qquad 
 B(\phi\to\gamma a_0\to\gamma\pi^0\eta^0)\simeq 10^{-5}\,;
\end{eqnarray}
iii) it is predicted
\begin{eqnarray}
\label{a0gammgammaf0gammagamma}
&&\Gamma(f_0\to\gamma\gamma)<0.13\cdot\Gamma(a_0\to\gamma\gamma)\,,
\end{eqnarray}
that is on the verge of conflict with the experiment, compare Eqs.
(\ref{ga0ggexp}) and (\ref{gf0ggexp});\\
iv) it is also predicted
\begin{eqnarray}
\label{a0rhof0phi}
&& B(J/\psi\to a_0(980)\rho)=(3/\lambda\approx 6)\cdot
B(J/\psi\to f_0(980)\phi)\,,
\end{eqnarray}
that has almost no chance, compare Eqs. (\ref{a0rho}) and (\ref{f0phi}).

Note that the $\lambda$ independent prediction
\begin{eqnarray}
\label{f0phi/f2'phia0rho/a2rho}
&& B(J/\psi\to f_0\phi)/B(J/\psi\to f_2'\phi)=
B(J/\psi\to a_0)\rho)/B(J/\psi\to a_2\rho)
\end{eqnarray}
is excluded by the central figure in
\begin{eqnarray}
\label{f0phi/f2'phi}
&& B(J/\psi\to f_0(980)\phi) /B(J/\psi\to f_2'(1525)\phi)= 0.4\pm 0.23\,,
\end{eqnarray}
obtained from Eqs. (\ref{f0phi}) and (\ref{f2'phi}), compare with Eq.
(\ref{a0rho/a2rho}). But, certainly, experimental error is too large.
Even twofold increase in accuracy of measurement of Eq. (\ref{f0phi/f2'phi})
could be crucial in the fate of the scenario under discussion.

The prospects to consider the $f_0(980)$-meson as the near $s\bar s$-state
(\ref{f0ss}) and the $a_0(980)$-meson as the four-quark state (\ref{a0qqqq})
with the coincidental mass degeneration is rather gloomy especially as the
four-quark model with the symbolic structure
\begin{eqnarray}
\label{f0qqqq}
&& f_0=s\bar s(u\bar u+d\bar d)\cos\theta /\sqrt{2}+
u\bar ud\bar d\sin\theta\,,
\end{eqnarray}
built around the MIT-bag \cite{jaffe-77}, justifies all unusual features of 
the $f_0(980)$-meson \cite{achasov-84,achasov-9188,achasov-98}.

As for the molecular model, wherein the $a_0(980)$ and $f_0(980)$-mesons are 
the extended bound states of the $K\bar K$-system \cite{isgur}, it seems that
the experiment leave no chance to this model now \cite{achasov-98}.

As for the traditional question, where are the scalar two-quark states from 
the lowest P-wave multiplet with the quark structures (\ref{a0qq}) and 
(\ref{f0qq}), there is no a tragedy with it now. All members of this 
multiplet are established \cite{pdg-98,achasov-98}.

\section{The theoretical grounds for the four-quark model}

A few words on the theoretical grounds for the four-quark nature of the 
$f_0$ and $a_0$ mesons. It was shown  in the context of the MIT-bag
\cite{jaffe-77} that the low-lying scalar four-quark nonet as bound state of
diquarks ($T_a=\varepsilon_{abc}\bar q^b\bar q^c$ and $\bar T^a=
\varepsilon^{abc}q_bq_c$, note that similar diquarks binding up with quarks
to form the baryon octet) arises from the strong binding energy in such a
configuration due to a hyperfine interaction Hamiltonian of the form
\begin{eqnarray}
&&H_{hf}=-\Delta\sum\vec s_i\cdot\vec s_j\vec F_i\cdot\vec F_j\,,
\quad\vec s=\vec\sigma/2\,,\ \ \vec F=\vec\lambda/2\,.
\nonumber
\end{eqnarray}

In the last few years the true renaissance has been going in treatments of
$\pi\pi$ and $\pi K$ scattering with help of phenomenological linear $\sigma$
models, see, for example,
\cite{achasov-94,schechter-95,tornqvist-95,scadron-95,ishida-96}.
It has been argued on occasion that the corresponding scalar mesons are
quark-antiquark states. But in fact at the Lagrangian level there is no
difference in the formulation of the two-quark and four-quark cases
\cite{schechter-99}.

\section{Conclusion}

So, there are many reasons to consider the $a_0(980)$ and $f_0(980)$ mesons
as the four-quark states. Nevertheless, in summary one emphasizes once again
that the further study of the decays $\phi\to\gamma f_0$ and 
$\phi\to\gamma a_0$; $J/\psi\to a_0\rho$, $f_0\omega$, $f_0\phi$, 
$a_2\rho$, $f_2\omega$, and $f_2'\phi $; $a_0\to\gamma\gamma$ and 
$f_0\to\gamma\gamma$; $D_s\to f_0\pi$ and $D_s\to a_0\pi$ \cite{lipkin-99} 
will enable one to solve the question on the $a_0(980)$ and $f_0(980)$-meson 
nature, at any case to close the above scenarios.

\end{document}